\documentclass[aps,prb,showpacs,superscriptaddress,floatfix]{revtex4}
\usepackage{graphicx}
\usepackage{amssymb}
\usepackage{amsmath}
\usepackage{bm}
\usepackage{xcolor}
\usepackage[utf8]{inputenc}
\usepackage[english]{babel}
\newcommand{\br}{{\bm r}}
\newcommand{\brp}{{\bm r'}}
\newcommand{\bx}{{\bm x}}
\newcommand{\by}{{\bm y}}
\newcommand{\bxp}{{\bm x'}}
\newcommand{\byp}{{\bm y'}}
\newcommand{\bo}{{\bm (}}
\newcommand{\bc}{{\bm )}}
\newcommand{\ua}{\uparrow}
\newcommand{\da}{\downarrow}
\newcommand{\eg}{{e.g.,}\ }
\newcommand{\ie}{{i.e.,}\ }
\newcommand{\dx}{d^3x}
\newcommand{\dy}{d^3y}
\newcommand{\dr}{d^3r}
\newcommand{\drp}{d^3r'}

\newcommand{\apots}{\Delta}
\newcommand{\apot}{\Delta}

\newcommand{\apoh}{{{w^{(s)}}}}
\newcommand{\apohp}{{{w^{(s)'}}}}
\newcommand{\adenss}{\chi^{}}
\newcommand{\adens}{\chi^{}}

\renewcommand{\r}{({\bf r})}
\newcommand{\rp}{({\bf r}')}
\newcommand{\rrp}{({\bf r},{\bf r}')}

\newcommand{\rs}{\rho_s}

\newcommand{\eme}{\ell}
\newcommand{\la}{\langle}
\newcommand{\ra}{\rangle}
\newcommand{\be}{\begin{equation}}
\newcommand{\ee}{\end{equation}}
\newcommand{\di}[4]{}

\begin{document}
\title{Density-functional description of superconducting and magnetic 
proximity effects across a tunneling barrier}

\author{Jorge~Quintanilla}
\affiliation{ISIS facility, STFC Rutherford Appleton Laboratory,
 Harwell Science and Innovation Centre, 
Didcot OX11 0QX, U.K.}
\author{Klaus~Capelle}
\author{Luiz N.~Oliveira}
\affiliation{Departamento de F\'{\i}sica e Inform\'atica, Instituto de
F\'{\i}sica de S\~ao Carlos, Universidade de S\~ao Paulo, Caixa
Postal 369, 13560-970 S\~ao Carlos, S\~ao Paulo, Brazil}
\date{\today}
\begin{abstract}
A density-functional formalism for superconductivity {\em and}
magnetism is presented. The resulting relations unify previously
derived Kohn-Sham equations for superconductors and for non-collinear
magnetism. The formalism, which discriminates Cooper pair
singlets from triplets, is applied to two quantum liquids coupled by
tunneling through a barrier. An exact expression is derived, relating
the eigenstates and eigenvalues of the Kohn-Sham equations, unperturbed
by tunneling, on one side of the barrier to the proximity-induced
ordering potential on the other.
\end{abstract}
                                                                                
\pacs{71.15.Mb, 74.20.Rp, 74.45.+c, 74.78.Fk, 75.70.Cn}
%\color{gray}
\maketitle

%\tableofcontents

\section{Introduction\label{intro}}
Condensed-matter phenomena owe much of their variety to the
multifarious properties of inhomogeneous electron liquids. Long-range
order, gaps, screening, enhanced correlations, anomalies,
resonances\textemdash a multiplicity of effects results from the
diversity of chemical compositions and crystal structures. Not
surprisingly, density-functional theory (DFT), a programme dedicated
to describing the effect of atomic-scale inhomogeneity on the electron
liquid, has acquired mounting
prominence.\cite{Koh99:1253,DreizlerGross90:} Its success in the
description of properties associated with the microscopic
inhomogeneities due to lattice potentials provides the foundation upon
which {\em ab initio} procedures are built. To describe ordered
states, specialized formalisms comprising other order parameters along
with the density have been
proposed.\cite{BaH72:1629,GuL76:4274,JG89:689,OGK88:2430,CaO00:15228}
 
Macroscopic inhomogeneities have received less attention from the DFT
community. In particular a physical boundary separating distinct order
parameters is a nesting ground for concepts and applications, of which
the discovery of giant magnetoresistances is the most persuasive
example.\cite{BBFp88_2472,BGSp89_4828} More recently, advances in
fabrication techniques and experimental probes (notably X-ray and
neutron reflectometry) have revealed that the behaviour at such
boundaries can be quite unexpected. To give one example, novel
magnetic order has been observed in the superconducting side of a
ferromagnet-superconductor interface
[Ref.~\onlinecite{2006-Chakhalian-et-al}; see also
  Ref.~\onlinecite{2006-Santamaria} and references therein]. Yet one
finds in the literature no DFT broadly applicable to junctions.

To fill this void, we present here a density-functional theory of
superconducting and magnetic materials. A generalization of previously
published theories for the individual
orderings,\cite{BaH72:1629,GuL76:4274,JG89:689,OGK88:2430,CaO00:15228}
the new formalism is designed to describe long-range order with
coexisting charge, magnetic and superconducting order
parameters. Competition or co-existence of different forms of order is
known to occur in bulk high-temperature,\cite{htsc} heavy fermion,
\cite{heavyferm1,heavyferm2} and organic \cite{organic}
superconductors as well as manganites.\cite{manganites} Here we will
focus on proximity effects in junctions between differently ordered
quantum liquids.\cite{foot:prox2003}

While studies of superconductivity in a specific material can be
restricted to Cooper pairs of given spin, either
singlets\cite{OGK88:2430,CG97:325} or triplets,\cite{CG97:325} the
more general setups that we target require parallel treatment of the
singlet pair, the triplet pair, and the magnetization densities. A
triplet superconductor coupled to a non-collinear antiferromagnet is, \eg
 well within the scope of our formalism.

As a general application, we consider two quantum liquids with
different order parameters separated by a thin barrier that allows
tunneling. Under these circunstances, we demonstrate that the
Kohn-Sham (KS) equations yielding the ground-state energies and densities
on one side of the barrier can be decoupled from the analogous
equations for the energies and densities on the opposite side. In each
decoupled KS Hamiltonian, an effective potential obtained from
the solution of the unperturbed KS equations, i.~e., the
KS equations in the absence of tunneling, represents the
opposite side. This potential gives mathematical substance, within DFT, to the
proximity effect. In a normal metal-superconductor junction, the
decoupled KS Hamiltonian adds correlation to the Bogolubov-de
Gennes equations.\cite{deG69a,DEG64:225} In a normal
metal-antiferromagnet junction, it generates analogous equations
describing the staggered proximity field induced on the nonmagnetic
side.

Our presentation starts out with a cursory review of DFT for
superconductivity and for magnetism. Section~\ref{sec:competing}
presents the formalism for coexisting order parameters. The resulting
KS equations are applied in Section \ref{sec:tunnel} to a barrier
separating two quantum liquids, and the effective proximity potential
is derived. In Section~\ref{sec:gener-magn-interf} the formalism is
generalised to magnetic interfaces. Finally,
Section~\ref{sec:conclusion} lists our conclusions.

\section{Density-functional theory for ordered quantum
  liquids\label{sec:dft}} The following two subsections briefly
recapitulate the DFT approach to ordered quantum liquids. They
highlight those aspects of the theory that will prove particularly
important in our formulation and others that were brought to view in recent
publications, posterior to the original references.

\subsection{Density-functional theory for superconductors\label{sec:scintro}}
Superconductors still pose a challenge to electronic-structure
theorists. Progress in that area had a late start. Long after
band-structure calculations based on DFT had provided valuable
information about such normal-state properties as Fermi surface
geometries, single-particle spectra, and electronic densities of
states, well after DFT had yielded such lattice properties related to
superconductivity as phonon dispersion relations, the superconducting
state still lay outside the realm of {\em ab initio} electronic-structure
calculations.

Model Hamiltonians had provided much of what was known about
superconductivity. The reduced BCS Hamiltonian,\cite{BCS57:1175} the
Hubbard Hamiltonian\cite{Hub63:238} and its variations, and the
Bogolubov-de-Gennes mean-field equations\cite{DEG64:225,deG69a} are
examples, none of which seemed adaptable to a density-functional
formulation. In view of its reliance on empirical information on phonon
spectra and Coulomb matrix elements, not even the detailed microscopic
description of strong-coupling superconductivity in Eliashberg's
theory could be seamed to {\em ab initio} DFT.\cite{Eli60:696}

To circumvent such difficulties, an alternative approach was
proposed.\cite{OGK88:2430} Instead of functionals of the density, one
now studied functionals of two variables:
the normal density $n\r=\langle \hat{\Psi}^\dagger_\ua\r
\hat{\Psi}_\da\r \rangle$ and the superconducting order parameter
$\adens\rrp =\langle \hat{\Psi}_\ua\r \hat{\Psi}_\da\rp \rangle$.
In the same way that $\rho$ is coupled to an electric potential, or
the magnetization density of spin-DFT (SDFT) is coupled to a magnetic
field, the anomalous density $\adens$ was coupled to a pair
potential. In the same way that the KS equations generalize the
Hartree mean-field equations, extended KS equations were
derived that generalize the mean-field Bogolubov-de Gennes equations.

Two are the potentials in these KS Bogolubov-de Gennes
equations: an effective electric potential
\begin{equation}
  v_s\r = v_{ext}\r + v_H\r + v_{xc}\r,
\end{equation}
and an effective pair potential
\begin{equation}
\label{eq:5}
  \apots\rrp = \apot_{ext}\rrp + \apot_H\rrp + \apot_{xc}\rrp.
\end{equation}

If the (microscopic) inhomogeneity is due exclusively to the lattice
potential $v_{ext}$, only an infinitesimal external potential
$\apot_{ext}$ is needed, to break the gauge symmetry that would
otherwise annul the anomalous density in the self-consistent
cycle. (As pointed out in Ref.~\onlinecite{OGK88:2430}, however, and
further discussed in Section~\ref{sec:one-side}, macroscopic
inhomogeneities may generate non-infinitesimal external anomalous
potentials.) While the Hartree potential $v_H$ tends to make the
charge distribution uniform, the anomalous Hartree potential
$\apot_H$, the interaction of the anomalous density with its own pair
potential, tends to enhance the superconducting order parameter.  As
usual, the exchange-correlation potential $v_{xc}\r = \delta
E_{xc}[n,\adens] /\delta n\r$ is the derivative of the universal $xc$
functional of DFT with respect to the normal density.  Similarly,
$\apot_{xc}\rrp= \delta E_{xc}[n,\adens] / \delta \adens\rrp$
represents the exchange-correlation correction to the mean-field
approximation.  Among the generalizations of this formalism, we
mention one that will assist our analysis: the extension to triplet
superconductors.\cite{CG97:325}

The exchange-correlation functional $E_{xc}[n,\adens]$ is, of course,
unknown, and the DFT programme calls for first-principles
approximations. That at least in the context of phonon-mediated
superconductivity this programme can be followed to its end was
demonstrated by Gross and collaborators,
\cite{KML+99:2628,MLL+05:024546,FPL+05:037004,
  LML+05:024545,PFL+06:047003,FSM+07:054508,SPF+07:020511} who
constructed a functional with no adjustable parameters and applied it
to a number of materials. We note in passing that their breakthrough
has led to the first truly microscopic theory of conventional
superconductivity.

\subsection{Density-functional for magnetic systems\label{sec:staggintro}}

The fundamental variables in the standard formulation of DFT,
collinear spin-DFT (SDFT), are the spin-resolved densities
$n_\ua\r$ and $n_\da\r$. From these, the charge density $n\r=n_\ua\r
+ n_\da\r$ and the magnetization $m_z\r = \mu(n_\ua\r - n_\da\r)$ are
promptly recovered ($\mu_0$ is the Bohr magneton). We concentrate our
discussion on two features of magnetic systems that are foreign to
SDFT in the local spin-density approximation: non-collinearity and nonlocality.

To describe non-collinear magnetic structures, one may substitute the
magnetization vector ${\bm m}\r$ for its $z$-component. Elegant extensions of
non-collinear SDFT and succesful implementations have been cast in this
forge.\cite{San98:91,NS96:4420,KHS+88b,KHS+88:3482} However, while
local approximations suffice to describe the nearly uniform average
local magnetization in a ferromagnet, such approximations cannot be
expected to fully reproduce the strong {\em nonlocal} correlations
associated with spin waves in a non-collinear antiferromagnet.\cite{2007-Gidopoulos}

Faced with this difficulty, one might construct nonlocal, \eg
orbital-dependent, functionals of the local variables $n\r$ and ${\bf
  m}\r$. \cite{2007-Sharma-et-al} A simpler alternative is suggested
by the procedure that, starting with spin-independent DFT, constructed
SDFT and DFT for superconductors: a density sensitive to the
ground-state correlations characteristic of the phenomenon under study
is added to the list of fundamental variables. To the set
$\{n_\ua\r,n_\da\r\}$ of SDFT variables, non-collinear
antiferromagnetism thus adds the {\em staggered
  density}\cite{CaO00:376,CaO00:15228,CSO01:1017}
\begin{equation}
  \rho_s\rrp=\langle \hat{\Psi}^\dagger_\ua\r \hat{\Psi}_\da\rp \rangle.
\end{equation}
The formal similarity between $\rho_s$ and the anomalous density in
DFT for superconductors is not accidental: already in 1958 it was
recognized that the restricted particle-hole
transformation\cite{And58:1900,FeM66:245} 
\begin{eqnarray*}
  \hat{\Psi}_\ua\r &\mapsto& \hat{\Psi}_\ua\r \\ 
  \hat{\Psi}_\da\r &\mapsto& \hat{\Psi}^\dagger_\da\r  
\end{eqnarray*}
converts the BCS procedure into a mean-field theory of antiferromagnetism.

Central to the development of staggered DFT is a coupling between the staggered
density and a staggered potential $S\rrp$, a nonlocal generalization
of the magnetic field ${\bf B\r}$. In the resulting KS
equations, the staggered density is coupled to the effective staggered
potential
\begin{equation}
  S_s\rrp =
  S_{ext}\rrp + S_H\rrp + S_{xc}\rrp.
  \label{effectiveS}
\end{equation}
Exploring the analogy with the external pair potential in DFT for
superconductivity, Refs.~\onlinecite{CaO00:15228} and
\onlinecite{CaO00:376} conjectured that, more than a mathematical
artifact, the first term on the right-hand side could be interpreted
as a proximity effect, a potential induced near a
non-collinear antiferromagnet.

Two features distinguish staggered DFT from collinear SDFT. (i) The
diagonal element (${\bf r'}={\bf r}$) of the staggered density
determines the $x$- and $y$-components of the magnetization: $m_x\r =
\mu_0[\rs({\bf r},{\bf r}) + \rs^*({\bf r},{\bf r})]$ and $m_y\r = i
\mu_0[\rs^*({\bf r},{\bf r}) - \rs({\bf r},{\bf r})]$. This
restriction yields a formalism equivalent to local non-collinear SDFT.
\cite{San98:91,NS96:4420,KHS+88b,KHS+88:3482}
Even under this restriction, even for perfectly collinear states,
non-collinear DFT is more powerful than standard (i.~e., collinear)
SDFT. In the latter, the specification of the quantization direction
breaks rotational symmetry. The staggered density restores that
symmetry: collinear SDFT is blind to, \eg  a magnetization along the
$x$ axis, since $n_\ua= n_\da =0$. Staggered SDFT, by contrast,
extracts the $x$ and $y$ components of the magnetization from
$\rho_s$.

(ii) The nonlocal dependence $\rho_s=\rho_s\rrp$ enhances the
superiority of staggered DFT over SDFT, and makes it more powerful
than local non-collinear SDFT.  To see this, it is sufficient to
recall that the electron liquid possesses an instability against the
formation of spin-density waves
(SDW).\cite{herring64:_magnet,Ove60:462,Ove62:1437} The driving force
of the Overhauser instability is the {\em staggered Hartree interaction}
\begin{equation}
\label{eq:6}
  U_x[\rho_s] = -\int d^3r \int
  d^3r' \frac{|\rho_s\rrp|^2}{|{\bf r}-{\bf r}'|}.
\end{equation}
This interaction, which formally appears as a Hartree term involving the 
staggered density, arises from evaluating the exchange diagram with
two-component spinors. It tends to push the energy of the SDW state below
that of the paramagnetic state.\cite{CaO00:376,CaO00:15228}

Standard formulations of SDFT miss this energy reduction altogether;
sophisticated approximations to the exchange-correlation functional
are needed to account for it.\cite{2007-Sharma-et-al} By contrast, the
right-hand side of Eq.~(\ref{eq:6}) is trivially incorporated in the
definition of the staggered DFT exchange-correlation energy
functional; the epithet ``staggered Hartree energy'' emphasizes its
formal similarity to the usual Hartree term while the subscript $x$ is
a reminder of its physical origin in the exchange
diagram.\cite{CaO00:376,CaO00:15228,CSO01:1017} Its contribution
$S_H\rrp = \delta U_x/\delta \rs\rrp$ to the effective Hartree
potential $S_s\rrp$ defines the second term on the right-hand side of
Eq.~(\ref{effectiveS}). Staggered DFT has been shown to recover both
the Overhauser instability in the exchange-only approximation and its
suppression by correlation.\cite{CaO00:15228}

\section{Density-functional formalism for competing order
  parameters\label{sec:competing}} We now turn to a formulation that
extends the results recapitulated in Section \ref{sec:dft} to more
complex ground states, with coexisting or competing order parameters.
Our derivation following closely Ref.~\onlinecite{CG97:325}
for the superconducting order parameter and
Ref.~\onlinecite{CaO00:15228} for the non-collinear order 
parameter, we list but the key equations.  

\subsection{Many-body Hamiltonian\label{sec:many-body-hamilt}}

 We consider an interacting Hamiltonian that includes external potentials
 coupled to all the physical observables of interest:
\begin{eqnarray}
  \hat{H}&=&\hat{T}+\hat{U}+\sum_{\sigma}\int \dr\,
  \hat{n}_{\sigma}\left(\br\right)v_{\sigma}\left(\br\right)
  \nonumber \\
  &&+\int \dx\,\dy \bo\hat{\rho}_{S}(\bx,\by)S(\bx,\by)
  -\hat{\adenss}(\bx,\by) \apots^*(\bx,\by)-
    \sum_{m=-1}^1\hat{\adens}_{m}(\bx,\by)
    \apot^*_{m}(\bx,\by)+\text{H.~c.}\bc.
\label{eq:1}
  \label{eq:1}
\end{eqnarray}
Here, the density operators
$\hat{n}_{\sigma}(\br)\equiv\hat{\psi}_{\sigma}^{\dagger}(\br)\hat{\psi}_{\sigma}(\br)$,
$\hat{\rho}_{s}(\br,\brp)\equiv \psi^\dagger_\ua(\br)\psi_\da(\brp)$,
$\hat{\adenss}(\br,\brp)\equiv [\psi(\br)\psi(\brp)]_s$ (the singlet
combination of the two field operators), and
$\hat{\adens}_{m}(\br,\brp)\equiv
[\psi(\br)\psi(\brp)]_m$ (the $m=-1,0$ and $1$ triplet
combinations of the field operators) are coupled to the external
potentials $v_{\sigma}\left(\br\right)=
v\left(\br\right)+\sigma\mu_{B}B_{z}\left(\br\right)$, $S(\bx,\by)$
(the staggered potential \cite{CaO00:15228}) $\apots(\bx,\by)=
\apots(\by,\bx)$ (the singlet pairing potential \cite{CG97:325}), and
$\apot_{m}(\bx,\by)= -\apot_{m}(\by,\bx)$ (the $m$ component of the triplet
pairing potential\cite{CG97:325}), respectively. As usual, the kinetic
energy operator is $\hat{T}\equiv
\sum_{\sigma=\uparrow\,\downarrow}\int \dr\,
\psi_{\sigma}^{\dagger}\left(\br\right)\,
\left(-\hbar^{2}\nabla^{2}/2m\right)\,\psi_{\sigma}\left(\br\right)$
where $m$ is electronic mass, and as in
Refs.~\onlinecite{OGK88:2430,CG97:325}, we assume the following form for the
electron-electron interaction:
\begin{equation}
  \hat{U}:=\frac{1}{2}\sum_{\beta_{1},\alpha_{1},\alpha_{2},\beta_{2}}
  \int \dy_{1}\dx_{1}\dx_{2}\dy_{2}
  \hat{\psi}_{\beta_{1}}^{\dagger}\left(\by_{1}\right)
  \hat{\psi}_{\alpha_{1}}^{\dagger}(\bx_{1})\,
  U\left(\by_{1},\beta_{1};\bx_{1},\alpha_{1};
\bx_{2},\alpha_{2};\by_{2},\beta_{2}\right)\,\hat{\psi}_{\alpha_{2}}
(\bx_{2})\hat{\psi}_{\beta_{2}}(\by_{2})\label{U},
\end{equation}
where the electron-electron interaction comprises the Coulomb
repulsion and a nonlocal spin-dependent term, needed to represent,
\eg the phonon-mediated attraction in conventional superconductors or
a spin-fluctuation mediated interaction in unconventional
ones: 
\begin{equation}
  U\left(\by_{1},\beta_{1};\bx_{1},\alpha_{1};
  \bx_{2},\alpha_{2};\by_{2},\beta_{2}\right)=
  \frac{q^{2}}{\left|\bx_{1}-\by_{1}\right|}
  \delta(\bx_{1}-\bx_{2})\delta
  \left(\by_{1}-\by_{2}\right)
  \delta_{\alpha_{1},\alpha_{2}}\delta_{\beta_{1},\beta_{2}}
  +W\left(\by_{1},\beta_{1};\bx_{1},\alpha_{1};
  \bx_{2},\alpha_{2};\by_{2},\beta_{2}\right)\label{general U}.
\end{equation}
Since the following arguments make only implicit reference to $\hat W$, to be
concise we have chosen a non-retarded potential to illustrate Eq.~(\ref{general U}).

Lest the reader be puzzled by the asymmetry in our treatment of the
pair density, which is resolved into a triplet and a singlet
components, while the staggered density is not, we note that the
restricted particle-hole transformation $\psi_\uparrow(\br)\mapsto
\psi_\uparrow(\br)$, $\psi_\downarrow(\br)\mapsto
\psi^\dagger_\uparrow(\br)$ would turn spins into isospins. Under this
transformation, the Cooper pairs would comprise an isospin dublet
coupled to an anomalous magnetic-field like potential, while the
staggered density would decompose into an isospin singlet and an
isospin triplet. Although the two approaches are mathematically
equivalent, we find the language of spins more attractive for the
present purposes than that of isospins.

\subsection{Kohn-Sham equations\label{sec:equations}}

Given $\hat U$, the ground-state energy is a functional of the
densities $n_{\sigma}$ ($\sigma=\ua,\da$), $\rho_s$, $\adens$, and
$\adens_m$ ($m=-1,0,1$), which can be written as
\begin{eqnarray}
E[n_\ua,n_\da,\rho_s,\adens,\adens_{m=0},\adens_{m=+1},\adens_{m=-1}]=
F_{HK}[n_\ua,n_\da,\rho_s,\adens,\adens_{m=0},\adens_{m=+1},\adens_{m=-1}]+
\nonumber \\
\sum_{\sigma}\int \dr\,
n_{\sigma}\left(\br\right)v_{\sigma}\left(\br\right)
%\nonumber \\ &&
+\int \dx\,\dy \bo\rho_{S}(\bx,\by)S(\bx,\by)
-\adenss(\bx,\by) \apots^*(\bx,\by)-
\sum_{m=-1}^1\adens_{m}(\bx,\by)
\apot^*_{m}(\bx,\by)+\text{c.c.}\bc,
\label{eq9}
\end{eqnarray}
where the potential energy in the various external potentials has been written
explicitly, and the kinetic and interaction energy are combined into the 
Hohenberg-Kohn internal-energy functional $F_{HK}$:
\be
F_{HK} = \la \hat{T} \ra + \la \hat{U} \ra = T_s + U_{MF} + E_{xc}.
\ee
In the last equation we defined the exchange-correlation functional 
$E_{xc}=\la \hat{T} \ra - T_s + \la \hat{U} \ra - U_{MF}$ in terms of
the kinetic energy $T_s$ of a noninteracting system with densities
$n_\ua,n_\da,\rho_s,\adens,\adens_{m=0},\adens_{m=+1},\adens_{m=-1}$, and
the mean-field approximation $U_{MF}$ to the full interaction energy 
$\la \hat{U} \ra$. Formally, this mean-field approximation is given by
\begin{eqnarray}
U_{MF}=
\frac{1}{2}\sum_{\beta_{1},\alpha_{1},\alpha_{2},\beta_{2}}
  \int \dy_{1}\dx_{1}\dx_{2}\dy_{2}\,
  \la \hat{\psi}_{\beta_{1}}^{\dagger}\left(\by_{1}\right)
  \hat{\psi}_{\alpha_{1}}^{\dagger}(\bx_{1})\ra\,
  U\left(\by_{1},\beta_{1};\bx_{1},\alpha_{1};
\bx_{2},\alpha_{2};\by_{2},\beta_{2}\right)\,
\la\hat{\psi}_{\alpha_{2}} (\bx_{2})\hat{\psi}_{\beta_{2}}(\by_{2})\ra
\nonumber \\
+
\frac{1}{2}\sum_{\beta_{1},\alpha_{1},\alpha_{2},\beta_{2}}
  \int \dy_{1}\dx_{1}\dx_{2}\dy_{2}\,
  \la \hat{\psi}_{\beta_{1}}^{\dagger}\left(\by_{1}\right)
  \hat{\psi}_{\beta_{2}}(\by_{1})\ra\,
  U\left(\by_{1},\beta_{1};\bx_{1},\alpha_{1};
\bx_{2},\alpha_{2};\by_{2},\beta_{2}\right)\,
\la\hat{\psi}_{\alpha_{1}}^\dagger(\bx_{1})
\hat{\psi}_{\alpha_{2}}(\bx_{2})\ra
\nonumber \\
-
\frac{1}{2}\sum_{\beta_{1},\alpha_{1},\alpha_{2},\beta_{2}}
  \int \dy_{1}\dx_{1}\dx_{2}\dy_{2}\,
  \la \hat{\psi}_{\beta_{1}}^{\dagger}\left(\by_{1}\right)
  \hat{\psi}_{\alpha_{2}}(\bx_{2})\ra\,
  U\left(\by_{1},\beta_{1};\bx_{1},\alpha_{1};
\bx_{2},\alpha_{2};\by_{2},\beta_{2}\right)\,
\la\hat{\psi}_{\alpha_{1}}^\dagger(\bx_{1})\hat{\psi}_{\beta_{2}}(\by_{2})\ra
\label{UMF},
\end{eqnarray}
Some contributions to $U_{MF}$ have appeared before in more restricted 
formulations, such as the conventional Hartree term
\be
E_H=\frac{1}{2} \int d^3x \int d^3y\, \frac{n({\bf x}) n({\bf y})}{|{\bf x}-{\bf y}|},
\ee
the anomalous Hartree term
\be
E_H^a=\int d^3x \int d^3y\, \frac{|\adens({\bf x},{\bf y})|^2}{|{\bf x}-{\bf y}|},
\ee
and the staggered Hartree term
\be
E_H^s = -\int d^3x \int d^3y\, \frac{|\rho_s({\bf x},{\bf y})|^2}{|{\bf x}-{\bf y}|} \equiv U_x[\rho_s].
\ee 
We stress that while $E_H^a$ is repulsive, and thus detrimental to 
superconductivity,\cite{KML+99:2628} $U_x[\rho_s]$ is attractive, and favours
formation of spin-density waves.\cite{CaO00:376,CaO00:15228,CSO01:1017}

The KS procedure, applied to this total-energy functional, leads to 
single-particle equations for the four-component spinor
\begin{equation}\label{spinor}
  \Phi^{\ell}\left(\br\right)=
\left(u_{\uparrow}^{\ell}\left(\br\right),u_{\downarrow}^{\ell}\left(\br\right),
v_{\uparrow}^{\ell}\left(\br\right),v_{\downarrow}^{\ell}\left(\br\right)\right)^{T},
\end{equation}
with particle (hole) wavefunctions
$u_{\sigma}({\bm r})$ [$v_{\sigma}({\bm r})$] for each electron spin,
$\sigma=\uparrow,\downarrow$. The symbol $T$ indicates transposition,
while the superscript $\ell$ labels the self-consistent solutions of the KS
equations, the first of which is
\begin{equation}
  H^{(s)}\left[\apoh\right]\Phi^{\ell}
  =\varepsilon_{\ell}\Phi^{\ell},
  \label{KS eq}
\end{equation}
where $\apoh$ denotes the set of the seven potentials
$\{v_{\sigma}^{(s)},S^{(s)},\apots^{(s)},\apot_{m}^{(s)}\}$
$(\sigma=\ua,\da, m=-1,0,1)$. The KS Hamiltonian is 
\begin{equation}
H^{(s)}=\left(\begin{array}{cccc}
h+v_{\uparrow,s} & \hat{S}^{(s)} & -2\hat{\apot}_{1}^{(s)} &
{-\hat{\apots}^{(s)}-\hat{\apot}_0^{(s)}}\\
\hat{S}^{(s)\dagger} & h+v_{\downarrow,s} &
{\hat{\apots}^{(s)}-\hat{\apot}_0^{(s)}} &
-2\hat{\apot}^{(s)}_{-1}\\
2\hat{\apot}_{1}^{(s)*} & {\hat{\apot}_0^{(s)*}+\hat{\apots}^{(s)*}} 
& -h-v_{\uparrow,s} & -\hat{S}^{(s)*}\\
{\hat{\apot}_0^{(s)*}-\hat{\apots}^{(s)*}} & 2\hat{\apot}_{-1}^{(s)*} & -\hat{S}^{(s)T} 
& -h-v_{\downarrow,s}\end{array}\right),
\label{Capelle-Oliveira-01-Eq16}
\end{equation}
with the shorthands $h$ for $-\hbar^{2}\nabla^{2}/2m$, and
$\hat{S}^{(s)}$, $\hat{\apots}^{(s)}$ and $\hat{\apot}_m^{(s)}$ for the
integral operators associated with the non-local potentials
$S^{(s)}\left(\br,\brp\right)$, $\apots^{(s)}\left(\br,\brp\right)$
and $\apot_m^{(s)}\left(\br,\brp\right)$ by
\begin{subequations}
  \begin{eqnarray}
    \hat{S}^{(s)}:\, f(\bx) & \mapsto & \int
    \dy S^{(s)}(\bx,\by)f(\by), \label{DEF S int op}\\
    \hat{\apots}^{(s)}:\, f(\bx) & \mapsto & \int
    \dy \apots^{(s)}(\bx,\by)f(\by)
    \label{DEF Delta int op}\qquad\text{and}\\
    \hat{\apot}^{(s)}_{m}:\, f(\bx) & \mapsto & \int
    \dy \apot^{(s)}_{m}(\bx,\by)f(\by)
    \label{DEF Delta_sig int op}.
  \end{eqnarray}
\end{subequations}
The effective single-body potentials are defined by
\begin{subequations}
  \begin{eqnarray}
    v_{\sigma s} & = &
    v_{\sigma}+\frac{\delta U_{MF}}{\delta n_{\sigma}}
+ \frac{\delta E_{xc}}{\delta n_{\sigma}},\label{sc_v}\\
    S^{(s)} & = &
    S+\frac{\delta U_{MF}}{\delta\rho_s}
+\frac{\delta E_{xc}}{\delta\rho_s}
    \label{sc S}\\
    \apot^{(s)} & = &
    \apot+\frac{\delta U_{MF}}{\delta\adenss}
+\frac{\delta E_{xc}}{\delta\adenss},
    \qquad\text{and}
    \label{sc Delta}\\
    \apot_{m}^{(s)} & = &
    \apot_{m}+\frac{\delta U_{MF}}{\delta\adens_{m}}
+\frac{\delta E_{xc}}{\delta\adens_{m}}.
    \label{sc Delta_m}
  \end{eqnarray}
\end{subequations}

The derivation of Eq.~(\ref{Capelle-Oliveira-01-Eq16}) from Eq.~(\ref{eq9}) 
is analogous to that of the simpler KS Hamiltonians in 
Refs.~\onlinecite{OGK88:2430} and \onlinecite{CaO00:15228}.
First, we note that minimization of the interacting Hamiltonian of 
Eq.~(\ref{eq:1}) with respect to $n_{\sigma}$ results in the Euler equation
\begin{equation}
\frac{\delta T_s}{\delta n_{\sigma}\r} + \frac{\delta U_{MF}}{\delta n_{\sigma}\r} 
+\frac{\delta E_{xc}}{\delta n_{\sigma}\r} + v_{\sigma}\r = 0,
\end{equation}
and similar equations for the minimization with respect to the other
densities. 

In a second step, we construct a non-interacting many-body system
with Hamiltonian $\hat{H}_{s}$ obtained by substracting $\hat{U}$ from the
interacting Hamiltonian of Eq.~(\ref{eq:1}) and replacing the external 
potentials $v_{\sigma}$, $S$, $\Delta$ and $\Delta_m$ with $v_{\sigma,s}$, 
$S^{(s)}$, $\Delta^{(s)}$ and $\Delta_m^{(s)}$, respectively. These 
self-consistent potentials are chosen such that the non-interacting system
has the same densities $n_{\uparrow}$, $n_{\downarrow}$, $\rho_s$, $\chi$
and $\chi_m$ as given by the original Hamiltonian, $\hat{H}$. Minimization 
of $\hat{H}_{s}$ with respect to $n_{\sigma}$ leads to the Euler equation
\begin{equation}
\frac{\delta T_s}{\delta n_{\sigma}\r} + v_{\sigma,s}\r = 0. 
\end{equation}

The two Euler equations can be combined to yield Eq.~(\ref{sc_v}). In the 
same manner, minimization with respect to the other
densities yields Eqs.~(\ref{sc S},\ref{sc Delta},\ref{sc
Delta_m}). To solve the effective problem, we need to diagonalise
$\hat{H}_s$. We achieve this by a a linear (Bogolubov)
transformation from the electron creation and annihilation
operators, $\hat{\psi}_{\sigma}^{\dagger}({\bf
r})$ and $\hat{\psi}_{\sigma}({\bf r})$, to new fermionic creation and
annihilation operators $\hat{\gamma}_l^{\dagger},\hat{\gamma}_l$:
\begin{equation}
\hat{\psi}_\sigma({\bf r}) = \sum_l \left[ u_{\sigma}^l({\bf r})
\hat{\gamma}_l + v_{\sigma}^l({\bf r})^* \hat{\gamma}_l^{\dagger}
\right]. 
\end{equation}
This transformation renders the effective single-particle Hamiltonian in 
diagonal form: $\hat{H}_s = \mbox{constant}+\sum_l \epsilon_l
\hat{\gamma}_l^{\dagger}\hat{\gamma}_l$. The single-particle energies
$\varepsilon_l$ are the eigenvalues of the KS Hamiltonian 
(\ref{Capelle-Oliveira-01-Eq16}), while the coefficients $u_{\sigma}^l$ 
and $v_{\sigma}^l$ of the Bogolubov transformation are the components of 
the spinors $\Phi^l$.

In terms of these coefficients, the ground-state densities are given by
\begin{subequations}
  \begin{eqnarray}\label{eq:14}
    n_{\sigma}\left(\br\right) & = &
    \sum_{\ell:\varepsilon_{\ell}<0}\left|u_{\sigma}^{\ell}\left(\br\right)\right|^{2}
    +\sum_{\ell:\varepsilon_{\ell}>0}\left|v_{\sigma}^{\ell}\left(\br\right)\right|^{2}
    \qquad(\sigma=\uparrow,\downarrow),\\ 
    \rho_s(\bx,\by) & = &
    \sum_{\ell:\varepsilon_{\ell}<0}u_{\uparrow}^{\ell*}(\bx)u_{\downarrow}^{\ell}(\by)
    +\sum_{\ell:\varepsilon_{\ell}>0}v_{\uparrow}^{\ell}(\bx)v_{\downarrow}^{\ell*}(\by),
    \label{eq:15}\\
    \adenss(\bx,\by) & = &
    \sum_{\ell:\varepsilon_{\ell}<0}[v^{\ell*}(\bx)u^{\ell}(\by)]_s
    +\sum_{\ell:\varepsilon_{\ell}>0}[u^{\ell}(\bx)v^{\ell*}(\by)]_s,\qquad
    \text{and}\label{eq:16}\\
    \adens_{m}(\bx,\by) & = &
    \sum_{\ell:\varepsilon_{\ell}<0}[v^{\ell*}(\bx)u^{\ell}(\by)]_m
    +\sum_{\ell:\varepsilon_{\ell}>0}[u^{\ell}(\bx)v^{\ell*}(\by)]_m.
    \label{eq:17}
  \end{eqnarray} 
\end{subequations}
The ground-state energy is then obtained by computing the functional
$E[n_\ua,n_\da,\rho_s,\adens,\adens_{m=0},\adens_{m=+1},\adens_{m=-1}]$.
%\begin{eqnarray}
%E\left[\adenh\right]
% &\equiv& 
%F_{HK}\left[\adenh\right]+\sum_{\sigma}
%\int\dr\,\,n_{\sigma}\left(\br\right)v_{\sigma}\left(\br\right)\nonumber\\&&+
%\int \dx\,\,\dy\,\,\left[\rho_s(\bx,\by)S(\bx,\by)
%-\adenss(\bx,\by)
%\apot^*(\bx,\by)
%-\sum_m\adens_{m}(\bx,\by)
%\apot^*_{m}(\bx,\by)+\text{c.~c.}\right],
%\end{eqnarray} 
%where $\adenh$ denotes the set of the seven densities
%$\{n_{\sigma},\rho_s,\adenss, \adens_m\}$ $(\sigma=\ua,\da,
%m=-1,0,1)$, and the universal Hohenberg-Kohn (HK) functional
%$F_{HK}\left[\adenh\right]\equiv\left\langle \hat{T}+\hat{U}\right\rangle$

%Here, $T^{(s)}\left[\adenh\right]$ is the HK functional for $\hat{U}=0$.

The KS Eqs.~(\ref{KS eq}-\ref{sc Delta_m}) determine the ground-state energy
and densities for external potentials ranging from microscopically to
the macroscopically inhomogeneous, as well as from periodic to
disordered. Our interest in the proximity effect focuses the
following discussion on the inhomogeneities associated with a
one-dimensional discontinuity separating two semi-infinite regions.

\section{Specialization to interfaces\label{sec:tunnel}}
\begin{figure}
\begin{centering}\includegraphics[width=0.35\textwidth]{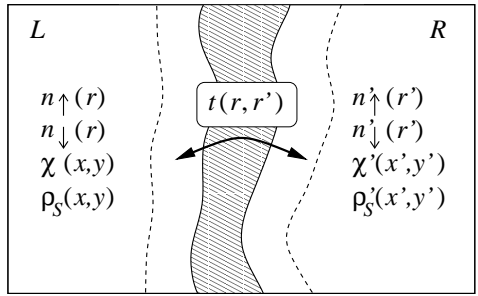}\par\end{centering}
\caption{\label{FIG-physical situation}The physical system under consideration
(see text): the shaded area represents the physically inaccessible
region across which tunneling can take place, while the dashed lines
indicate the extent of the $L$ and $R$ regions on which the tunneling
amplitude $t\left(\br,\brp\right)$ is important. }
\end{figure}

We consider two semi-infinite spatial regions, labelled $L$ (left) and
$R$ (right), separated by a potential barrier that is wide in
comparison with the range of the interaction $\hat U$ in either $L$ or
$R$. Though high, the barrier is finite and hence allows electronic
tunneling between $L$ and $R$. As Fig.~\ref{FIG-physical situation}
indicates, we denote the position in $L$ ($R$) by $\br$ ($\brp$). To
save space, we will indicate the densities and potentials on the $R$
semi-space by a prime. Thus, \eg $v(\br \in L)$ [$v(\br \in R)]$ will
be denoted $v$ ($v'$).

Our model Hamiltonian reads
\begin{equation}
\hat{H}=\hat{H}_{L}+\hat{H}_{T}+\hat{H}_{R}\label{3-piece hamiltonian}.
\end{equation}
Here, $\hat{H}_L$ ($\hat{H}_R$) are of the form $\hat H+\hat V$, where
$\hat H$ is defined in Eq.~(\ref{eq:1}), and $\hat V$
is an infinite barrier centered at the interface between $L$ ($R$) and
the tunneling region, which enforces orthogonality between the
eigenfunctions of $\hat H_L$ and $\hat H_R$.\cite{Pra63:1083} The
phenomenological tunneling Hamiltonian
\begin{equation}\label{eq:2}
\hat{H}_{T}=\sum_{\sigma}\int_{L}\dr\int_{R}\drp\,
\left[\hat{\psi}_{\sigma}^{\dagger}\left(\br\right)\, 
t\left(\br,\brp\right)\,\hat{\psi}_{\sigma}
\left(\brp\right)+H.c.\right]
\end{equation}
provides transport across the barrier.

%\subsection{Model of the tunneling barrier\label{sec:model-tunnel}}

We complete the phenomenlogical description of the interface by assuming
%[{\bf Jorge, please clarify:} why "assuming"? Can't this KS equation be derived 
%by a Bogolubov transformation from the Hamiltonian containing the tunneling 
%term?] 
that the only effect of the tunneling term $\hat{H}_{T}$ in the Hamiltonian
(\ref{3-piece hamiltonian}) is to make the KS equations become

\begin{equation}
  \left(\begin{array}{cc}
    H^{(s)}\left[\apoh\right] & H_{T}^{(s)}\\
    H_{T}^{(s)\dagger} &  H^{(s)}\left[\apohp\right]
  \end{array}\right)
  \left(\begin{array}{c}  
    \Phi_{L}^{\nu}\\
    \Phi_{R}^{\nu}\end{array}\right)
  =\varepsilon_{\nu}\left(\begin{array}{c}
    \Phi_{L}^{\nu}\\
    \Phi_{R}^{\nu}\end{array}\right),\label{KS with tunnelling}
\end{equation}
where the tunneling KS Hamiltonian $H^{(s)}_{T}$ is defined by 

\begin{equation}\label{eq:3}
H_{T}^{(s)}=\left(\begin{array}{cccc}
\hat{t} & 0 & 0 & 0\\
0 & \hat{t} & 0 & 0\\
0 & 0 & -\hat{t}^{*} & 0\\
0 & 0 & 0 & -\hat{t}^{*}\end{array}\right).
\end{equation}
The integral operator $\hat{t}$ is given by 
$\hat{t}:\, f\left(\br\right)\mapsto\int \drp\,
t\left(\br,\brp\right)f\left(\brp\right)$.

Our phenomenological description of the barrier contrasts with the
microscopic treatment of the semi-infinite regions. Instead of
deriving and solving Kohn-Sham equations for the shaded region in
Fig.~\ref{FIG-physical situation}, we prefer to follow Prange's
prescription \cite{Pra63:1083}, even if we have to rely on
phenomenological considerations to define the tunneling matrix
$t\left({\bf r},{\bf r}'\right)$. This simplification focuses our
analysis on the interaction between the $L$ and $R$ regions.  Under
appropriate circumstances, as shown below, it leads to an independent
set of Kohn-Sham equations for each region, a result combining
mathematical convenience with physical appeal. Under such
circumstances, to determine the densities in one of the regions, we
can rely on first-principles calculations in which the external
potential captures phenomenologically the influence of the material
across the barrier.

\subsection{Decoupled Hamiltonians\label{sec:one-side}}
The KS equations (\ref{KS with tunnelling}) describe the electronic
state of $L$ and $R$. Let us now derive a set of KS equations that
describe, formally, only the states of $R$, but take into account the
influence on such states of the charge, spin and superconducting order
that are potentially present in $L$.

For $\br\in L$, since the eigenfunctions of the KS Hamiltonian with $\hat{H}_{T}=0$
form a complete set, we can write
\begin{equation}
\Phi_{L}^{\nu}=\sum_{\ell}\left(\tilde{\Phi}_{L}^{\ell\dagger},
\Phi_{L}^{\nu}\right)\tilde{\Phi}_{L}^{\ell}\label{phi_L in terms of phi_tilde_L},
\end{equation}
where the tildes denote the absence of tunneling, and the scalar
product $\left(\tilde{\Phi}_{L}^{\ell\dagger}, \Phi_{L}^{\nu}\right)$
comprises integration over the spatial variable and summation over
spinor indices. 

Insertion into the second row of Eq.~(\ref{KS with tunnelling}) then yields

\begin{equation}
\sum_{\ell}\left(\tilde{\Phi}_{L}^{\ell\dagger},\Phi_{L}^{\nu}\right)
H_{T}^{(s)}\tilde{\Phi}_{L}^{\ell}+H^{(s)}
\left[\apohp\right]
\Phi_{R}^{\nu}=\varepsilon_{\nu}\Phi_{R}^{\nu}.
\label{KS w tunnelling 2nd row with tildes} 
\end{equation}
 To eliminate $\Phi_{L}^{\nu}$ from this equation, we first write
 the eigenvalue equation defining $\tilde{\Phi}_{L}^{\ell}$, i.~e., the
 first row of Eq.~(\ref{KS with tunnelling}) with $H_{T}^{(s)}=0$,
 \begin{equation}
    H^{(s)}\left[\apoh\right]\tilde{\Phi}_{L}^{\ell}=
\tilde\varepsilon_{L}^{\ell}\tilde{\Phi}_{L}^{\ell},
 \end{equation}
and multiply it on the left by $\Phi_{L}^{\nu\dagger}$. Hermitean
conjugation then yields
\begin{equation}\label{eq:4}
\tilde{\Phi}_{L}^{\ell\dagger}
H^{(s)}\left[\apoh\right]
\Phi_{L}^{\nu}=\tilde{\varepsilon}_{L}^{\ell}\tilde{\Phi}_{L}^{\ell\dagger}\Phi_{L}^{\nu}
\end{equation}

Next, we multiply the first row of Eq.~(\ref{KS with tunnelling}) by
$\tilde{\Phi}_{L}^{\ell\dagger}$ on the left:
\begin{equation}
  \tilde{\Phi}_{L}^{\ell\dagger}H_{T}^{(s)}\Phi_{R}^{\nu}+\tilde{\Phi}_{L}^{\ell\dagger}
H^{(s)}\left[\apoh\right]
\Phi_{L}^{\nu}=\varepsilon_{\nu}\tilde{\Phi}_{L}^{\ell\dagger}\Phi_{L}^{\nu}
\end{equation}
We then subtract Eq.~(\ref{eq:4}) from this result, solve
for $\tilde{\Phi}_{L}^{\ell\dagger}\Phi_{L}^{\nu}$, and integrate over
the spatial variable to find that
\begin{equation}
  \left(\tilde{\Phi}_{L}^{\ell\dagger},\Phi_{L}^{\nu}\right)
  =\frac{1}{\varepsilon_{\nu}-\tilde{\varepsilon}_{L}^{\ell}}
  \left(\tilde{\Phi}_{L}^{\ell\dagger},H_{T}^{(s)}\Phi_{R}^{\nu}\right).
\end{equation}
We can now substitute the right-hand side for the scalar product on
the left-hand side of Eq.~(\ref{KS w tunnelling 2nd row with tildes})
and exploit the completeness of the wavefunctions $\tilde\phi_L^{\ell}$
($l=1,2,\ldots$) to define the effective external Hamiltonian
\begin{equation}
H_{ext}\left(\varepsilon_{\nu}\right)=
\sum_{\ell}H_{T}^{(s)}\tilde{\Phi}_{L}^{\ell}
\frac{1}{\varepsilon_{\nu}-\tilde{\varepsilon}_{L}^{\ell}}
\tilde{\Phi}_{L}^{\ell\dagger}H_{T}^{(s)},
\label{H_ext}
\end{equation}
so that Eq.~(\ref{KS w tunnelling 2nd row with tildes}) defines a
KS eigenvalue equation for $\brp \in R$:
\begin{equation}
{\bm(}
H^{(s)}\left[\apohp\right]
+H_{ext}\left(\varepsilon_{\nu}\right){\bm )} 
\Phi_{R}^{\nu}=\varepsilon_{\nu}\Phi_{R}^{\nu}
\label{KS w tunnelling 2nd row decoupled}
\end{equation}

\begin{figure}
\begin{centering}
\includegraphics[width=0.35\textwidth]{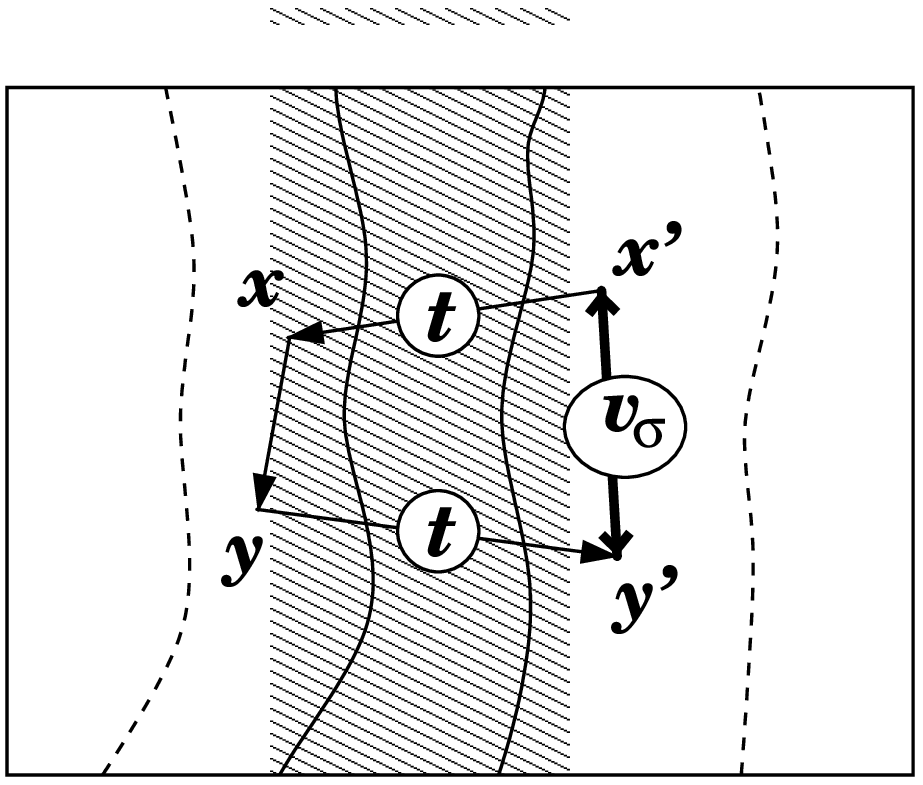}\par
\end{centering}
\caption{\label{FIG-tunnelling} The external non-local {}``potential''
of Eq.~(\ref{v_ext_b}) induced in the $R$ region describes new coupling
between points $\bxp$ and $\byp$ that takes into account
processes in which an electron would tunnel across te barrier, propagate
inside the differently-ordered electron liquid in the $L$ region,
and then tunnel back into $R$ (see text).}
\end{figure}

More explicitly, given that, if
$\left(\tilde{u}_{\uparrow},\tilde{u}_{\downarrow},\tilde{v}_{\uparrow},
\tilde{v}_{\downarrow}\right)^{T}$
is an eigenvector, with eigenvalue $\tilde{\varepsilon}$, then
$\left(\tilde{v}_{\uparrow},\tilde{v}_{\downarrow},
\tilde{u}_{\uparrow},\tilde{u}_{\downarrow}\right)^{\dagger}$
is also an eigenvector, with eigenvalue $-\tilde{\varepsilon}$, we can
cast the effective external potential in the form
\begin{equation}
H_{ext}\left(\varepsilon_{\nu}\right)=\left(\begin{array}{cccc}
\hat{v}^{(s)}_{\uparrow}\left(\varepsilon_{\nu}\right) 
& \hat{S}^{(s)}\left(\varepsilon_{\nu}\right) 
& -2\hat{\apot}^{(s)}_{1}\left(\varepsilon_{\nu}\right) 
&-\bo\hat{\apots}^{(s)}+\hat{\apot}_0^{(s)}\bc\left(\varepsilon_{\nu}\right)\\
\hat{S}^{(s)\dagger}\left(\varepsilon_{\nu}\right) 
&\hat{v}^{(s)}_{\downarrow}\left(\varepsilon_{\nu}\right) 
&\bo\hat{\apots}^{(s)}-\hat{\apot}_0^{(s)}\bc\left(\varepsilon_{\nu}\right) 
&-2\hat{\apot}^{(s)}_{-1}\left(\varepsilon_{\nu}\right)\\
2\hat{\apot}^{(s)*}_{1}\left(-\varepsilon_{\nu}\right) 
&\bo\hat{\apot}_0^{(s)*}+\hat{\apots}^{(s)*}\bc\left(-\varepsilon_{\nu}\right) 
&-\hat{v}_{\uparrow}^{(s)}\left(-\varepsilon_{\nu}\right) 
&-\hat{S}^{(s)*}\left(-\varepsilon_{\nu}\right)\\
\bo\hat{\apot}_0^{(s)*}-\hat{\apots}^{(s)*}\bc\left(-\varepsilon_{\nu}\right) 
&2\hat{\apot}_{-1}^{(s)*}\left(-\varepsilon_{\nu}\right) 
&-\hat{S}^{(s)T}\left(-\varepsilon_{\nu}\right) 
&-\hat{v}_{\downarrow}^{(s)}\left(-\varepsilon_{\nu}\right)\end{array}\right)
\label{H_ext explicit}
\end{equation}
where the integral operators $\hat{v}_{\sigma}\left(\varepsilon_{\nu}\right)$,
$\hat{S}\left(\varepsilon_{\nu}\right)$, $\hat{\apots}\left(\varepsilon_{\nu}\right)$,
and $\hat{\apot}_{m}\left(\varepsilon_{\nu}\right)$ ($m=-1,0,1$) have
the following kernels:
\begin{subequations}
  \begin{eqnarray}
    v_{\sigma}\left(\varepsilon_{\nu};\bxp,\byp\right) &=&
    \int_{L}\dx\,\, 
    t(\bxp,\bx)\,\,\,
    \sum_{\ell}\int_{L}\dy\,\frac{\tilde{u}_{L,\sigma}^{\ell}
      (\bx)\tilde{u}_{L,\sigma}^{\ell*}
      (\by)}{\varepsilon_{\nu}-\tilde{\varepsilon}_{L,l}}\,\,\, 
    t\left(\by,\byp\right),\label{v_ext_b}\\
    S\left(\varepsilon_{\nu};\bxp,\byp\right) &=
    &\int_{L}\dx\,\, t
    (\bxp,\bx)\,\,\,
    \sum_{\ell}\int_{L}\dy\,\frac{\tilde{u}_{L,\uparrow}^{\ell}
      (\bx)\tilde{u}_{L,\downarrow}^{\ell*}
      (\by)}{\varepsilon_{\nu}-\tilde{\varepsilon}_{L,l}}\,\,\, 
    t\left(\by,\byp\right),\label{S_ext_b}\\
    \apots\left(\varepsilon_{\nu};\bxp,\byp\right) &=
    &\int_{L}\dx\,\, 
    t(\bxp,\bx)\,\,\,
    \sum_{\ell}\int_{L}\dy\,\frac{[\tilde{u}_{L}^{\ell}(\bx)
        \tilde{v}_{L}^{\ell*}
        (\by)]_s}{\varepsilon_{\nu}-\tilde{\varepsilon}_{L,l}}\,\,\,
    (-1)t^{*}(\by,\byp)
    \label{Delta_ext_b},\quad\text{and}\\
    \apot_{m}\left(\varepsilon_{\nu};\bxp,\byp\right) &=
    &\int_{L}\dx\,\, 
    t(\bxp,\bx)\,\,\,
    \sum_{\ell}\int_{L}\dy\,\frac{[\tilde{u}_{L}^{\ell}
        (\bx)\tilde{v}_{L}^{\ell*}(\by)]_m}
        {\varepsilon_{\nu}-\tilde{\varepsilon}_{L,l}}\,\,\,
    \left(-1\right)t^{*}(\by,\byp).
    \label{Delta_sigma_ext_b}
  \end{eqnarray}
\end{subequations}
The effective external potential
$H_{ext}\left(\varepsilon_{\nu}\right)$ expresses mathematically the
proximity effect. The non-local, effective potential
$v_{\sigma}\left(\varepsilon_{\nu};\bxp,\byp\right)$ in
Eq.~(\ref{v_ext_b}) describes the virtual transition from point
$\bxp$ to \textbf{$\byp$} depicted in
Fig.~\ref{FIG-tunnelling}, with intermediate tunneling to point
$\bx$ and propagation to point $\by$. The right-hand
sides of Eqs.~(\ref{S_ext_b}-\ref{Delta_sigma_ext_b}) have analogous
interpretations, the propagation in $L$ now involving a spin flip
[Eq.~(\ref{S_ext_b})] or spin-conserving scattering into a hole
state [Eqs.~(\ref{Delta_ext_b})~and (\ref{Delta_sigma_ext_b})], respectively.

\subsection{General aspects of the decoupled Hamiltonian\label{sec:decoupled-ham}}

Equations~(\ref{v_ext_b}-\ref{Delta_sigma_ext_b}) decompose a pair of
quantum liquids, differently ordered and coupled by tunneling, into
two uncoupled liquids subject to effective, energy-dependent external
fields. The right-hand side of each equality depends only on
$\tilde{\Phi}_{L}^{\ell}$, not on $\Phi_{R}^{\nu}$ or on
$\Phi_{L}^{\nu}$. In words, the effective fields induced by the
proximity effect on one half-space depend only on the solutions of
the KS equations on the opposite side, unperturbed by
tunneling.

In general, the effective fields also depend on the energies
$\varepsilon_\nu$.  Whether it is easier to diagonalize energy
independent coupled Hamiltonians or energy-dependent uncoupled ones
then depends on specific aspects of the problem under study. If a
  low barrier separates degenerate states, the coupled-Hamiltonian
  formalism is natural and mathematically convenient. As pointed
out in Ref.~\onlinecite{OGK88:2430}, however, if the energy spectrum
$\tilde\epsilon$ on one of the sides is bounded by a gap, while that
on the other side is not, then the low KS eigenvalues
$\varepsilon_\nu$ can be neglected in the denominators of the sums on
the right-hand sides of Eqs.~(\ref{v_ext_b}-\ref{Delta_sigma_ext_b}),
and the potentials become energy independent. In this case, the
treatment of the decoupled Hamiltonians is clearly more efficient than
the solution of the coupled KS equations. From a conceptual viewpoint,
moreover, the emergence of proximity-induced effective potentials that
give rise to superconductivity or antiferromagnetism is appealing.

In the analysis of infinite systems, at the formal level, DFT starts
out by introducing a ficticious external potential, set equal to zero
after the KS equations are derived. Equation.~(\ref{eq:1}) is
an example. To solve the KS equations, at the operational
level, one likewise seeds the self-consistent loop with an artificial
external potential. In the vicinity of an ordered system, as
Eqs.~(\ref{v_ext_b}-\ref{Delta_sigma_ext_b}) show, this mathematical
expedient is neither formally nor operationally necessary. One can
therefore interpret the spontaneous ordering of an infinite system as
a residual consequence of a proximity effect. We see that this
interpretation, first offered in connection with superconductors in
Ref.~\onlinecite{OGK88:2430}, describes equally well magnetic,
superconducting and coexisting states.

In spite of the complex form on the right-hand side of Eq.~(\ref{H_ext
  explicit}), our attention to spin in the definition of the Cooper
pair densities guarantees that the KS equations couple only the
singlet (triplet) density to the singlet (triplet) potential. As the
analyses in subsections~\ref{sec:spec-case:-superc}~and
\ref{sec:spec-case:-magn} will show, if a conventional superconductor
is coupled to a non-collinear antiferromagnet by an interface through
which electrons can tunnel, the resulting proximity potential will
induce the formation of singlet Cooper pairs in the antiferromagnetic
material; likewise, on the superconducting side, the induced staggered
field scatters the Cooper pairs into other singlet states.

\subsection{Special case: superconductivity\label{sec:spec-case:-superc}}
This subsection and the next examine
Eqs.~(\ref{v_ext_b}-\ref{Delta_sigma_ext_b}) in two particularly
simple situations. Here, we detail the first arrangement, which
couples an unspecified material ($R$) to a singlet superconductor
($L$). We take the latter to be nonmagnetic, so that
$v_{s,\uparrow}-v_{s,\downarrow}=S_{s}=0$. 

To determine the unperturbed KS eigenstates in $L$, which contribute
to the right-hand sides of
Eqs.~(\ref{v_ext_b}-\ref{Delta_sigma_ext_b}),
we set $H_T^s=0$. A Bogolubov-Valatin transformation then diagonalizes the
matrix on the right-hand side of the resulting
Eq.~(\ref{Capelle-Oliveira-01-Eq16}).  The eigenspinors have the form
\begin{equation}
\label{eq:7}
  \tilde{\Phi}_{L}^{\alpha\eme}\r\equiv\left(\begin{array}{c}
    \tilde{u}_{L,\uparrow}^{\alpha \eme}\r\\
    \tilde{u}_{L,\downarrow}^{\alpha \eme}\r\\
    \tilde{v}_{L,\uparrow}^{\alpha \eme}\r\\
    \tilde{v}_{L,\downarrow}^{\alpha
      \eme}\left(\mathbf{r}\right)\end{array}\right)
  =\left(\begin{array}{c}
    u_{\eme}\r\\
    0\\
    0\\
    v_{\eme}\r\end{array}\right)\delta_{\alpha,1}+\left(\begin{array}{c}
    0\\
    u_{\eme}\r\\
    -v_{\eme}\r\\
    0\end{array}\right)\delta_{\alpha,2}\qquad(\alpha=1,2)
\end{equation}

The amplitudes $u_{\eme}\r$ and $v_{\eme}\r$ are solutions of the
Bogolubov-de Gennes equations [\ie Eq.~(7) in
  Ref.~\onlinecite{OGK88:2430}]. The invariance of the $L$ Hamiltonian
under the inversion $z\mapsto -z$ makes the KS eigenvalues independent
of the spin index $\alpha$, \ie $\tilde{\varepsilon}_{L}^{\alpha
  \eme}=\tilde\varepsilon_{\eme}$.  The eigenvalues moreover change
sign under the transformation
$\left(\tilde{u}_{\uparrow},\tilde{u}_{\downarrow},\tilde{v}_{\uparrow},
\tilde{v}_{\downarrow}\right)^{T}\mapsto
\left(\tilde{v}_{\uparrow},\tilde{v}_{\downarrow},
\tilde{u}_{\uparrow},\tilde{u}_{\downarrow}\right)^{\dagger}$. These
two symmetries considered, and the eigenfunctions on the right-hand
side of Eq.~(\ref{eq:7}) substituted on the right-hand sides of
Eqs.~(\ref{v_ext_b}-\ref{Delta_sigma_ext_b}), we find that
\begin{subequations}
  \begin{eqnarray}
    v_{\sigma}\left(\varepsilon_{\nu};\bx',\by'\right)  &=&
    \int_{L}d^3x\,\,
    t\left(\bx',\bx\right)\,\sum_{\eme}
    \int_{L}d^3y\,\,\frac{u_{\eme\sigma}\left(\bx\right)u_{\eme\sigma}^{*}
        \left(\by\right)}{\varepsilon_{\nu}-\tilde\varepsilon_{\eme}}\, 
    t\left(\by,\by'\right)\label{v_ext BCS},\\
    \apots\left(\varepsilon_{\nu};\bx',\by'\right) &=& 
    \int_{L}d^3{x}\,\,
    t\left(\bx',\bx\right)\,\sum_{\eme}
    \int_{L}d^3{y}\,\left[\frac{u_{\eme}\left(\bx\right)v_{\eme}^{*}
        \left(\by\right)}{\tilde\varepsilon_{\eme}-\varepsilon_{\nu}}\right]_s 
    t^{*}\left(\by,\by'\right)\label{Delta_ext BCS},\quad\text{and}\\
    S\left(\varepsilon_{\nu};\bx',\by'\right)  
    &=&  \apot_m\left(\varepsilon_{\nu};\bx',\by'\right)=0\qquad(m=-1,0,1).
    \label{Lambda_ext and S_ext BCS}
  \end{eqnarray}
\end{subequations}

Thus the superconductivity in $L$ induces only a normal and a singlet
pairing potential in $R$. When $R$ is a semi-infinite, normal metal,
we recover the SC-DFT result.~\cite{OGK88:2430}  Equation~(\ref{v_ext BCS}) is the
nonlocal normal potential alluded to, but not given explicitly, in
footnote 10 of Ref.~\onlinecite{OGK88:2430}.

\subsection{Special case: magnetism\label{sec:spec-case:-magn}}

As a second particular case, we consider a non-superconducting
material in $L$, \ie $\apot_{L}=0$. The eigenspinors of the
KS equations now have the form
\begin{equation}
  \tilde{\Phi}_{L}^{\tau m}\r\equiv\left(\begin{array}{c}
    \tilde{u}_{L,\uparrow}^{\tau m}\r\\
    \tilde{u}_{L,\downarrow}^{\tau m}\r\\
    \tilde{v}_{L,\uparrow}^{\tau m}\r\\
    \tilde{v}_{L,\downarrow}^{\tau m}\r\end{array}\right)=\left(\begin{array}{c}
    \varphi_{m,\uparrow}\r\\
    \varphi_{m,\downarrow}\r\\
    0\\
    0\end{array}\right)\delta_{\tau,p}+\left(\begin{array}{c}
      0\\
      0\\
      \varphi_{m,\uparrow}^{*}\r\\
      \varphi_{m,\downarrow}^{*}\r\end{array}\right)\delta_{\tau,h}\qquad(\tau=p, h),
\end{equation}
where the label $\tau=p$ and $\tau=h$ designates particle- and
hole-like KS quasiparticles, respectively. The amplitudes $\varphi_{m,\sigma}\r$
are solutions of an eigenvalue problem analogous to the Bogolubov-de Gennes
equations [Eq.~(15) in Ref.~\onlinecite{CaO00:15228}]. For a
given $m$, the corresponding eigenvalue $\varepsilon_{m}$ can either be
positive or negative, and the external potentials are
\begin{eqnarray}
v_{\sigma}\left(\varepsilon_{\nu};\bx',\by'\right) &=&
\int_{L}d^3{x}\,\,t\left(\bx',\bx\right)\,
\sum_{m}\int_{L}d^3{y}\,\,\frac{\varphi_{m,\sigma}\left(\bx\right)
\varphi_{m,\sigma}^{*}\left(\by\right)}{\varepsilon_{\nu}-\varepsilon_{m}}\,
t\left(\by,\by'\right)
\label{v_ext SDW}\\ 
S\left(\varepsilon_{\nu};\bx',\by'\right) &=&
\int_{L}d^3{x}\,\, t\left(\bx',\bx\right)\,
 \sum_{m}\int_{L}d^3{y}\,\,\frac{\varphi_{m,\uparrow}\left(\bx\right)
\varphi_{m,\downarrow}^{*}\left(\by\right)}{\varepsilon_{\nu}-\varepsilon_{m}}\,
t\left(\by,\by'\right)
\label{S_ext SDW}\\ 
\apots\left(\varepsilon_{\nu};\bx',\by'\right) &=&
\apot_{m}\left(\varepsilon_{\nu};\bx',\by'\right)=0\qquad(m=-1,0,1).
\label{Delta_ext and Lambda_ext SDW}
\end{eqnarray}
This set is analogous to Eqs.~(\ref{v_ext BCS}-\ref{Lambda_ext and
  S_ext BCS}) and gives substance to the image\cite{CaO00:15228} of the
external contribution to the staggered field in DFT as a proximity
effect, an interpretation that places additional emphasis on the analogy between
DFT for superconductors and for spin-density waves.

\color{black}\section{Generalization: magnetic
  interfaces\label{sec:gener-magn-interf}} The tunneling
Hamiltonian~(\ref{eq:3}) defines a inert barrier, one that merely
allows charge transport between regions $L$ and $R$. By contrast, this
section examines a magnetically active interface, in which spin-orbit
coupling or spin-flip scattering from impurities add off-diagonal
elements to the tunneling matrix. In special situations, a trivial
rotation of the quantization axis may be sufficient to diagonalize
that matrix over the entire barrier; in such instances, one expects
the barrier to induce spin polarization in both $L$ and $R$. 

More generally, however, the spatial dependence of the matrix elements
will bar global diagonalization, so that instead of simply redefining
the quantization axis, the barrier will turn into an inhomogeneous
source of spin flips. While the system, constituted by the $L$ and $R$
regions and the interface between them, is subject to global
conservation laws, the spin of the electrons in regions $R$ and $L$ is
no longer conserved. This section explores the consequences of that
rupture.  We shall see that while the potentials $v_{\sigma}$
($\sigma=\ua,\da$) and $S$ are only quantitatively affected, the
magnetic barrier breaks the independence between the singlet and the
triplet Cooper pairs. A magnetic interface separating a singlet
superconductor from a normal metal, for instance, will induce the
formation of triplet pairs in the latter.

More specifically, instead of Eq.~(\ref{eq:3}) we consider the
following tunneling KS Hamiltonian
\begin{equation}\label{eq:8}
  H_{Tmagn}^{(s)}=\left(\begin{array}{cccc}
    \hat{t}_{\ua\ua} & \hat{t}_{\ua\da} & 0 & 0\\
    \hat{t}_{\da\ua} & \hat{t}_{\da\da} & 0 & 0\\
    0 & 0 & -\hat{t}_{\ua\ua}^{*} & -\hat{t}_{\ua\da}^{*}\\
    0 & 0 & -\hat{t}_{\da\ua}^{*} & -\hat{t}_{\da\da}^{*}\end{array}\right),
\end{equation}
and follow the analysis leading from Eq.~(\ref{eq:3}) to
Eq.~(\ref{H_ext}), which now takes the form
\begin{equation}\label{eq:9}
H_{ext,magn}\left(\varepsilon_{\nu}\right)=
\sum_{\ell}H_{Tmagn}^{(s)}\tilde{\Phi}_{L}^{\ell}
\frac{1}{\varepsilon_{\nu}-\tilde{\varepsilon}_{L}^{\ell}}
\tilde{\Phi}_{L}^{\ell\dagger}H_{Tmagn}^{(s)}.
\end{equation}

From Eqs.~(\ref{spinor})~and (\ref{eq:8}), the computation of the
right-hand side of Eq.~(\ref{eq:9}) is straightforward, which brings
Eq.~(\ref{eq:9}) to the form of Eq.~(\ref{H_ext explicit}), with the kernels
\begin{subequations}
  \begin{eqnarray}
    v_{\tau}(\varepsilon_{\nu};\bxp,\byp) &=&
    \int_{L}\dx\,\,\,
        \sum_{\sigma,\sigma'}t_{\tau\sigma}(\bxp,\bx)\,\,\int_{L}\dy\,\,\,
        \sum_{\ell}\frac{\tilde{u}_{L,\sigma}^{\ell}
          (\bx)\tilde{u}_{L,\sigma'}^{\ell*}
          (\by)}{\varepsilon_{\nu}-\tilde{\varepsilon}_{L,l}}\,\,\, 
        t_{\sigma'\tau}(\by,\byp),\qquad(\tau=\ua,\da)
        \label{eq:10}\\
        S(\varepsilon_{\nu};\bxp,\byp) &=&
        \int_{L}\dx\,\,\,\sum_{\sigma,\sigma'}t_{\ua\sigma}(\bxp,\bx)\,\,\,\int_{L}\dy\,\,
        \sum_{\ell}\frac{\tilde{u}_{L,\sigma}^{\ell}
          (\bx)\tilde{u}_{L,\sigma'}^{\ell*}
          (\by)}{\varepsilon_{\nu}-\tilde{\varepsilon}_{L,l}}\,\,\, 
        t_{\sigma'\da}(\by,\byp),
        \label{eq:11}
  \end{eqnarray}
  and, with the notation $\mathcal{P}_{\tau\tau'}^{sm}$
  ($\tau,\tau'=\ua,\da$, $\mathcal{S}=0,1$, and
  $m=-\mathcal{S},\ldots,\mathcal{S}$) for the spin operator
  projecting the doublets $\tau$ and $\tau'$ onto their singlet
  ($\mathcal{S}=0$) or triplet ($\mathcal{S}=1$) combinations,\cite{Psinglet}
\begin{eqnarray}
    \Delta(\varepsilon_{\nu};\bxp,\byp)
    &=&\frac12\mathcal{P}_{\tau\tau'}^{00} \int_{L}\dx\,
    \sum_{\sigma,\sigma'}t_{\tau\sigma}(\bxp,\bx)\,\,\,
    \int_{L}\dy\,\,\,\sum_{\ell}\frac{\tilde{u}_{L,\sigma}^{\ell}
      (\bx)\tilde{v}_{L,\sigma'}^{\ell*}
      (\by)}{\varepsilon_{\nu}-\tilde{\varepsilon}_{L,l}}\,\,\,
    t^*_{\sigma'\tau'}(\by,\byp),
    \label{eq:12}\\
    \Delta_{m}(\varepsilon_{\nu};\bxp,\byp) &=&
    \frac12\mathcal{P}_{\tau\tau'}^{1m}
    \int_{L}\dx\,\,\,
    \sum_{\sigma,\sigma'}t_{\tau\sigma}(\bxp,\bx)\,\,\,\int_{L}\dy\,\,
    \sum_{\ell}\frac{\tilde{u}_{L,\sigma}^{\ell}
      (\bx)\tilde{v}_{L,\sigma'}^{\ell*}
          (\by)}{\varepsilon_{\nu}-\tilde{\varepsilon}_{L,l}}\,\,\, 
    t^*_{\sigma'\tau'}(\by,\byp)\qquad(m=0,\pm1).
    \label{eq:13}
\end{eqnarray}
\end{subequations}

While the projector $\mathcal{P}_{\tau\tau'}^{\mathcal{S}m}$ in
Eqs.~(\ref{eq:12})~and (\ref{eq:13}) ensures singlet and triplet
symmetry for $\Delta$ and $\Delta_m$, respectively, the sums over the
spin components $\sigma$ and $\sigma'$ involve all combinations of the
eigenvectors $\tilde u_{L\sigma}^\ell$ and $\tilde
v^{\ell*}_{L\sigma'}$, \ie contributions from both the singlet and the
triplet anomalous densities.  The product $\tilde
u_{L\ua}^\ell(\bx)\tilde v^{\ell*}_{L\ua}(\by)$, for instance, which
Eq.~(\ref{eq:17}) associates with $\adens_1(\by,\bx)$, contributes to
$\Delta$ and to all three components of $\Delta_m$; should a global
rotation of the quantization axis diagonalize both $t(\bxp,\bx)$ and
$t(\by,\byp)$, its contribution to the singlet potential $\Delta$
would vanish, so that the mismatch between the symmetries of the
densities and of the KS potentials would be restricted to the $z$
component of $\adens_m$ and $\apot_m$. In general, however, in the
presence of magnetically inhomogeneous interfaces neither the $z$
component nor the total spin on one side of the interface are
conserved.

In particular, if a singlet superconductor in region $L$ is coupled to a normal
metal in region $R$, the singlet anomalous density will contribute to
the right-hand side of Eq.~(\ref{eq:13}). We expect, therefore, 
triplet Cooper pairs to be proximity-induced in the normal metal,
along with singlet pairs due to the potential $\Delta$.

To conclude this section we note that the coupling between anomalous
densities with different symmetries is in line with spin
conservation. The off-diagonal terms in our phenomenological
Eq.~(\ref{eq:8}) arise from magnetic degrees of freedom in the
barrier, which interact with the spins of the electrons in $L$ and
$R$. Only the total spin $\bm{S}^2=(\bm{S}_L+\bm{S}_T+\bm{S}_R)^2$,
which includes the contribution $\bm{S_T}$ from such degrees of
freedom, must be conserved. While a tunneling matrix~(\ref{eq:8})
diagonalizable by a uniform rotation of its spin variables is
sufficient to conserve $S_L^2$ and $S_R^2$, and while at least in
special situations, this condition proves not
necessary,\cite{accidental} the right-hand side of Eq.~(\ref{eq:8})
will in general fail to commute with $S_R^2$ and hence allow
$\mathcal{S}\to\mathcal{S}\pm1$ Cooper-pair spin transitions.

%\color{gray}
\section{Conclusions\label{sec:conclusion}}
We have discussed a density-functional formalism describing
superconductivity and magnetism, a generalization encompassing DFT for
superconductors \cite{OGK88:2430,CG97:325} and DFT for antiferromagnets
\cite{CaO00:15228} that covers systems with coexisting order
parameters. The new formalism being particularly practical in
descriptions of two quantum liquids coupled by tunneling, we have
derived an exact expression relating the proximity field in region one
to the eigenfunctions and eigenvalues of the unperturbed KS
equations in the opposite region. 

The solution of the self-consistent cycle of equations presented in
Section \ref{sec:equations} gives individual access to the singlet and
the triplet Cooper pair densities, as well as to the staggered
magnetization. This feature of the formalism was explored in
Section~\ref{sec:decoupled-ham}, which demonstrated that the proximity
field induced by the singlet (triplet) density affects only the
singlet (triplet) pairs and showed that DFT opens attractive
perspectives for the study of the competition between order parameters
in junctions such as a conventional or unconventional superconductor
coupled to a non-collinear antiferromagnet. In
Section~\ref{sec:gener-magn-interf} we considered the special case of
a magnetic interface. We found that a \emph{singlet} superconductor
can in this case induce \emph{triplet} pairing in a normal metal. We
hope that our results stimulate further experimental scrutiny of
macroscopically inhomogeneous quantum liquids and serve as the basis
for an {\it ab initio} description of such systems.

 \acknowledgments This work received financial support from the FAPESP
 and CNPq. JQ thanks Nikitas Gidopoulos some useful comments. JQ
 acknowledges financial support from CCLRC (now STFC) in association
 with St. Catherine's College, Oxford.

%\bibliography{proximity,jquinta}

\begin{thebibliography}{46}
\bibitem[{acc()}]{accidental}
\bibinfo{note}{As an example of a tunneling matrix that cannot be diagonalized
  by a global rotation and still conserves $\bm{S}_L^2$ and $\bm{S}_R^2$,
  consider a matrix diagonalizable by a uniform rotation everywhere except in a
  region where the barrier is so high that the eigenfunctions $u^\ell$ and
  $v^\ell$ vanish.}
\end{thebibliography}
%
% bbl file put in by hand:

\end{document}